\documentclass[aps,pra,superscriptaddress,reprint,floatfix]{revtex4-1}
\usepackage{amstext,amssymb}
\usepackage{graphicx}
\usepackage{xcolor}
\usepackage{amsmath}
\usepackage{amsfonts}

\newcommand{\be}{\begin{equation}}
\newcommand{\ee}{\end{equation}}
\newcommand{\ben}{\begin{eqnarray}}
\newcommand{\een}{\end{eqnarray}}

\newcommand{\pare}[1]{\left( #1 \right)}

\newcommand{\ket}[1]{|#1\rangle}
\newcommand{\bra}[1]{\langle #1|}

\newcommand{\op}[2]{| #1\rangle \langle#2|}

\begin{document}
\title{Dynamical signatures of molecular symmetries in nonequilibrium quantum transport}
\author{Juzar Thingna}
%\email[]{juzar@smart.mit.edu}
\affiliation{Massachusetts Institute of Technology, Chemistry Department. Cambridge, Massachusetts 02139, USA}
\affiliation{Singapore-MIT Alliance for Research and Technology (SMART) Centre, Singapore 138602}
\author{Daniel Manzano}
\affiliation{Massachusetts Institute of Technology, Chemistry Department. Cambridge, Massachusetts 02139, USA}
\affiliation{Singapore University of Technology and Design, Engineering Product Development. 8 Somapah Road, Singapore 487372}
\affiliation{Universidad de Granada, Departamento de Electromagnetismo y F\'isica de la Materia and Instituto Carlos I de F\'isica Te\'orica y Computacional, 
Granada 18071, Spain}
\author{Jianshu Cao}
\email[]{jianshu@mit.edu}
\affiliation{Massachusetts Institute of Technology, Chemistry Department. Cambridge, Massachusetts 02139, USA}
\affiliation{Singapore-MIT Alliance for Research and Technology (SMART) Centre, Singapore 138602}
\date{\today}

\begin{abstract}
Symmetries play a crucial role in ubiquitous systems found in Nature. In this work, we propose an elegant approach to detect symmetries by measuring quantum currents. Our detection scheme relies on initiating the system in an anti-symmetric initial condition, with respect to the symmetric sites, and using a probe that acts like a local noise. Depending on the position of the probe the currents exhibit unique signatures such as a quasi-stationary plateau indicating the presence of meta-stability and multi-exponential decays in case of multiple symmetries. The signatures are sensitive to the probe and vanish completely when the timescale of the coherent system dynamics is much longer than the timescale of the probe. These results are demonstrated using a $4$-site model and an archetypal example of the para-benzene ring and are shown to be robust under a weak disorder.
\end{abstract}
\pacs{}
\maketitle

%---------------------introduction---------------
Symmetries are a vital ingredient to many molecular systems and play an important role in their transport properties \cite{Gross1996, Lehmann2002, Walschaers13, Denisov2014}. They lead to intriguing non-trivial consequences like the existence of multiple steady states \cite{Buca2012, Baumgartner2008} and could help design smart devices \cite{Manzano2014,Daniel2015}. Despite these advantages, the effect of symmetries in the transport properties of molecular systems have not been fully explored. Several studies on symmetric systems have explored the effects of dephasing \cite{Rai2010, Rai2011} and interference \cite{Solomon2010, Chen2014}.

A thorough knowledge of symmetries could lead to a better understanding of molecular junctions that have become promising candidates to build transport devices \cite{Tao2006, Aradhya2013}. They have been in the limelight due to the sizable quantum effects that have led to exciting effects in electronic \cite{Ventra2000, Thingna2014, Zhou2015}, heat \cite{ThingnaPRB2012, Lee2013, Wang2015, Segal16}, and excitonic \cite{Cao2009, Walschaers2013} transport. Despite the plethora of studies, transport signatures that arise solely due to the inherent molecular symmetry have not been put forward. These signatures could help identify the symmetric sites within the molecule thus providing a clear atomic picture of the system of interest. Using these signatures, we could explore molecular symmetry to design nanodevices with tunable transport properties. 

In this study we propose an efficient approach to detect symmetries in molecular junctions by measuring the excitonic currents. The main idea relies on using a probe that could act locally on each site \cite{Oteyza2013} thus breaking the symmetry of the system. Therefore, by varying the position of the probe and initiating the system in an appropriately chosen initial condition we find the excitonic currents vary from zero (for symmetric system) to a finite non-zero value (for systems with broken symmetry). The non-zero currents exhibit a long quasi-stationary plateau akin to a classical metastable state \cite{Darroch1967}. If there are multiple symmetries present within a system, like in the case of para-benzene, then the multiple symmetries can also be distinguished due to the presence of a multi-exponential decay in the current dynamics. The signatures we obtain are exclusive to symmetric systems and are found to be sensitive to the timescales of the system and the probe. Thus, we provide a comprehensive detection scheme of molecular symmetries and elucidate its dynamical effects.
\begin{figure*}
\includegraphics[width=\textwidth]{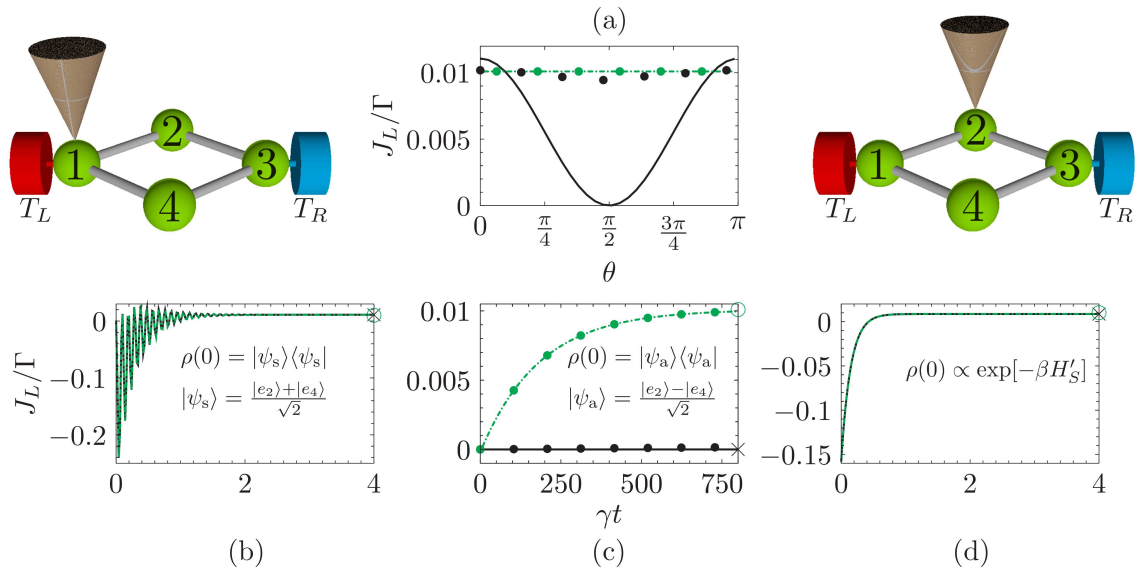}
\caption{\label{fig:figure1}\textbf{Effect of probe position on symmetry detection.} Panel a depicts the modulation of the steady-state excitonic currents in the $4$-site model as a function of the mixing angle, Eq.~(\ref{eq:mixing}), that determines the initial condition. Panels b, c, and d show the time evolution of the excitonic current with symmetric (panel b), antisymmetric (panel c), and canonical (panel d) initial conditions. The probe is positioned at sites $1$ (black solid line) and $2$ (green dashed-dotted line) for all panels. The black crosses and open green circles correspond to the nonequilibrium steady state values of the excitonic currents. The steady state values for panel b and d are small but non-zero. Note the different time unit in panel (c) indicating a very slow relaxation from the antisymmetric initial condition. The closed circles in panel (a) and (b) represent results for a weakly disordered system with disorder strength $\delta = 0.13$meV. The currents are averaged over 1000 samples of uniformly distributed random disorder [standard error of mean (SEM) is contained within each closed circle]. The black closed circles correspond to the probe being placed at site $1$, whereas the green closed circle correspond to the probe at site $2$. The system parameters are: $\varepsilon = -142.2$meV and $h =-9.35$meV. The lead and probe parameters are chosen as: $T_{L} = 330K$, $T_{R} = 270K$, $T = 300K$, $\Gamma = 196$GHz, $\gamma =19.6$GHz, $\omega_0 = 78.55$THz, and $\omega_{D}=1.96$ THz.}
\end{figure*}

\noindent \textbf{Results}\\
\noindent \textbf{Detecting symmetries.}
Symmetry detection forms the first crucial step to manipulate molecular junctions in order to build novel nanodevices, such as thermal switches \cite{Daniel2015}. One of the key properties of symmetric systems is the presence of multiple steady states. These occur when the reduced dynamics is confined to an invariant subspace that is not disturbed by the dissipative leads (see methods section for a general proof). Our proof shows that when the interaction of the reduced system doesn’t perturb the system out of the invariant subspace, the symmetries are preserved and reflected in the reduced dynamics in the form of multiple steady states. In other words irrespective of the coupling strength or the form of the master equation that governs the reduced dynamics a symmetric system will possess multiple steady states. 

Detecting such symmetries is highly non-trivial, especially in experimental set-ups \cite{Baykusheva2016}, and in this work we propose a scheme that would help achieve this objective and provide robust signatures in the transient currents. Our scheme consists of two steps: i) Initiating the molecular system in a dark state and ii) Using a probe that acts locally on the molecular sites to detect which sites are symmetric. The probe when placed on different sites could break or preserve the symmetries of the system. The symmetry breaking would produce a non-zero current whereas when symmetries are preserved the current would be zero (due to the dark state initial condition that belongs to an invariant subspace). 

For \emph{perfectly} symmetric systems the switching phenomenon can be observed even in the steady state. But perfect symmetry is difficult to obtain in experimental set-ups and most systems are subject to conformational disorder. Even for a `close to' symmetric system with a weak disorder (either static or dephasing) there would be a single unique steady state. Hence the steady-state currents will not switch depending on the probe position. In this case the dynamical regime could provide essential signatures. If the disorder is weak it would affect the dynamical currents on the longest timescale and thus during the intermediate times we expect that a weakly disordered system would \emph{mimic} a perfectly symmetric system. 

The breaking of symmetries could also affect the relaxation times and help unravel the multiplicity of the steady-state degeneracy. Thus, our goal of detecting symmetries could be achieved using the two step approach that could even help design a control over the timescale of the relaxation process. Inspired by these ideas and the timely application of detecting symmetries we explore the points presented in this section using two concrete examples of a $4$-site model and the archetypal benzene molecule.

\noindent \textbf{$4$-site model.} The minimal nonequilibrium model that exhibits molecular symmetries is a $4$-site model as sketched in Figure \ref{fig:figure1}. The Hamiltonian of the system is given by,
\begin{align}
\label{eq:H}
H_{{ S}} = \begin{pmatrix}
\varepsilon_g & 0 \\
0 & H_{{ S}}^{\prime}
\end{pmatrix},
\end{align}
where the $4$-site Hamiltonian $H_{{ S}}^{\prime}$ reads
\begin{align}
H_{{ S}}^{\prime} &= \varepsilon\sum_{i=1}^{4}\op{e_i}{e_i} + h\sum_{\left<i,j \right>} \op{e_i}{ e_j}.
\end{align}
In the above the sum over $\left<i,j \right>$ indicates nearest neighbor interactions with cyclic boundary conditions. This Hamiltonian within the H\"{u}ckle theory of atomic molecular orbitals can be interpreted as having 4 atomic sites with nearest neighbor interaction \cite{Streitwieser1961, Lowe1978}. The diagonal elements $\varepsilon$ represent the combination of kinetic energy and Coulomb integrals whereas the off-diagonal coefficient $h$ represents the nearest-neighbor overlap between atomic sites. 

The system is connected to left and right leads that govern the transitions of excitons between the ground state $e_g$ and the excited states $e_1$ and $e_3$. The presence of two leads at different temperatures creates a nonequilibrium setup that generates a flow of excitons within the molecular system. The effect of the leads is modeled via a Lindblad term in the master equation (see methods section) whose Lindblad operators read
\begin{align}	
\label{eq:Lindblad1}
A_{{ L}1}&=\op{e_g}{ e_1}, 
&A_{{ L}2} = \op{e_1}{ e_g},  \\
\label{eq:Lindblad2}
A_{{ R}1}&= \op{e_g}{e_3}, 
&A_{{ R}2}= \op{e_3}{e_g}. 
\end{align}
The nonequilibrium $4$-site model exhibits two nonequilibrium steady states \cite{Buca2012, Albert2014, Manzano2014} due to the existence of a unitary operator $\Pi=\exp\left(\ket{e_2}\bra{e_4}+\ket{e_4}\bra{e_2}\right)$. The operator $\Pi$, also known as the \emph{symmetry operator}, obeys $[\Pi,H_{ S}]=[\Pi,A_{\alpha k}]=0~(\alpha = L,R;~k=1,2)$ due to which the time evolution of a general density matrix $\rho$ can be decomposed into invariant subspaces (see methods section for a general proof). It is worth noting here that the Lindblad description provides a minimalist set-up and does not affect the overall results of this work.

Our goal in this work is to devise a scheme to detect molecular symmetries using the excitonic currents. In order to achieve this objective we introduce a probe that acts on the system as a site specific dynamical noise without injecting or extracting excitons. When the probe acts on sites $1$ ($S=\op{e_1}{e_1}$) or $3$ ($S=\op{e_3}{e_3}$) the Hamiltonian of the probe-system interaction commutes with the original symmetry operator (see methods section). On the other hand the action of the probe on site $2$ ($S=\op{e_2}{e_2}$) or $4$ ($S=\op{e_4}{e_4}$) breaks the symmetry of the system leading to a single \emph{unique} nonequilibrium steady state. We incorporate the effects of the probe via a Redfield tensor as shown in the methods section. Similar probes, e.g. the B\"{u}ttiker probe \cite{Buttiker1986}, have been used to investigate effects of coherence \cite{Buttiker1986}, interference \cite{Chen2014} and also to measure temperatures at a molecular level \cite{Dubi2011}.

The behavior of excitonic currents for the $4$-site model starting from different initial conditions and different probe positions is shown in Fig.~\ref{fig:figure1}. The probe is acting either on site $1$ (unbroken symmetry case) or on site $2$ (broken symmetry case). The three initial conditions correspond to the symmetric exchange of sites $2$ and $4$, $\rho_{\text{s}} = 0.5 \left( \op{e_2}{e_2} + \op{e_4}{e_4} + \op{e_2}{e_4} + \op{e_4}{e_2} \right)$, antisymmetric exchange, $\rho_{\text{a}} = 0.5 \left( \op{e_2}{e_2} + \op{e_4}{e_4} - \op{e_2}{e_4} - \op{e_4}{e_2} \right)$, or a canonical distribution $\rho_c\propto \exp\left[-\beta H_{S}^{\prime}\right]$. 

Figure \ref{fig:figure1}a shows the steady-state currents as a function of the mixing angle ($\theta$) between the symmetric and antisymmetric initial conditions. We set the initial condition 
\begin{eqnarray}
\label{eq:mixing}
\rho (0) & = \cos^{2}(\theta)\rho_{\text{s}} + \sin^{2}(\theta)\rho_{\text{a}}.
\end{eqnarray}
For the perfectly symmetric system when the probe is placed at site $1$ (unbroken symmetry case, black solid line) the excitonic currents can be controlled by tuning the mixing angle $\theta$. On the other hand when the probe is placed at site $2$ (broken symmetry case, green dashed-dotted line) we obtain a unique steady state and hence there is no modulation of the steady state currents due to the mixing angle. Thus, in case of perfectly symmetric systems the steady state currents exhibit a clear signature of the underlying symmetry in terms of the switching behavior depending on the probe position. Unfortunately, perfectly symmetric systems are rarely feasible and in order to describe a realistic system we introduce a static random disorder $\Delta\varepsilon_{i}$ at each site $e_{i}$  $\left( \forall i\neq g \right)$ chosen from a uniform distribution with width $\delta$ $\left( \Delta\varepsilon_i \in  [-\frac{\delta}{2},\frac{\delta}{2}] \right)$. The excitonic currents are averaged over 1000 samples (enough to achieve convergence) of disorder to obtain the closed circles in Fig. \ref{fig:figure1}a. Clearly, in this case the probe position (site $1 \equiv$ black closed circles and site $2\equiv$ green closed circles) plays no role since we have only one unique steady state. Therefore, in a realistic system with conformational disorder it is impossible to observe any steady-state signature of the underlying symmetry. Hence, this observation serves as our main motivation to study the \emph{dynamics} of the excitonic currents, due to the separation of time scale of the disorder and the symmetries, in order to observe the dynamical signatures related to symmetries.

\begin{figure}
\includegraphics[width=\columnwidth]{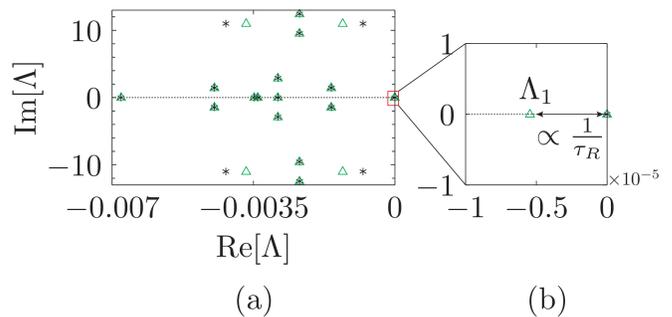}
\caption{\label{fig:figure2}\textbf{Dissipative Liouvillian spectrum for the $4$-site model.} (a) Eigenspectrum of the dissipative Liouvillian for the $4$-site model. The asterisk correspond to the eigenvalues $\Lambda$ when the probe is placed at site $1$, whereas the open green triangles correspond to probe positioned at site $2$. (b) shows a magnification around the zero eigenvalue. There are 2 degenerated eigenvalues at zero when the probe is positioned at site $1$. These eigenvalues split when the probe is at site $2$ and the resultant gap dictates the relaxation time $\tau_R$. The dashed black line marks the Im[$\Lambda$] = 0 axis. All other parameters are the same as in Fig.~\ref{fig:figure1}.}
\end{figure}
In order to achieve our objective we first study the transient currents in perfectly symmetric systems. We express the time evolution of the reduced density matrix in terms of an eigenfunction expansion \cite{Jung2002} of the dissipative Liouvillian $\mathcal{L}$ (see Eq. (\ref{eq:ME}) from methods section) as, 
\begin{align}
\label{eq:rhot}
|\rho(t)\rangle= \sum_{k} \op{\Phi^{r}_{k}}{\Phi^{l}_{k}}\rho(0)\rangle e^{\Lambda_{k}t},
\end{align}
where we are working in the Liouville-Fock space \cite{prosen:njp08}, mapping the density matrix to a vector and the Liouvillian $\mathcal{L}$ to a matrix $L$ that follows $L\ket{\Phi^{r}_k} = \Lambda_{k}\ket{\Phi_{k}^{r}}$ and $\bra{\Phi^{l}_k}L = \Lambda_{k}\bra{\Phi_{k}^{l}}$. Here $\bra{\Phi^{l}}$ and $\ket{\Phi^{r}}$ are the left and right eigenoperators of the dissipative non-Hermitian Liouvillian that form a dual basis, and $\Lambda_{k}$ are their corresponding eigenvalues. The dynamics then clearly depends on two important factors, namely, the eigenvalues $\Lambda_k$ and the weight determined by the overlap of the corresponding left eigenoperator with the initial density matrix $\bra{\Phi^{l}_k}\rho(0)\rangle$. The eigenspectrum for the dissipative Liouvillian for our $4$-site model is shown in Fig.~\ref{fig:figure2}. The probe, when placed at site $2$, produces a subtle effect by splitting the degenerated zero eigenvalues as indicated in Fig.~\ref{fig:figure2}b. The eigenvalue that separates from the zero (indicated by $\Lambda_{1}$) corresponds to the unstable steady state and has an antisymmetric structure. Thus, its left eigenoperator overlap with the symmetric initial condition is zero and the dynamics of the excitonic currents is governed by the remaining eigenvalues and eigenoperators that are not drastically affected due to the probe position. Therefore, we find that for the symmetric initial condition there are no dynamical signatures of symmetries in the excitonic currents as shown in Fig.~\ref{fig:figure1}b.

The antisymmetric initial condition (also known as the dark state) \cite{Lidar2003} is one of the multiple steady states (unstable manifold). In this state the excitonic currents are exactly zero as shown by the black solid line in Fig.~\ref{fig:figure1}c. As soon as we break the symmetry we observe a slow relaxation for the excitonic currents due to the transition from the unstable to the unique steady state. The timescale of this transition is now governed by the dissipative Liouvillian eigenvalue corresponding to the unstable steady state $\tau_{R} \propto \Lambda_{1}^{-1}$ which is several magnitudes larger than the relaxation time for the symmetric initial condition (note the $x$-axis in Figs.~\ref{fig:figure1}b and c). We also find that there is no change in the time-dependent excitonic currents if the probe is placed at site $2$ or $4$. Thus, the slow quasi-stationary relaxation, when initiating from a dark state, provides a clear signature of symmetries in the dynamical currents that inherently depend on the probe position.

\begin{figure*}
\includegraphics[width=\textwidth]{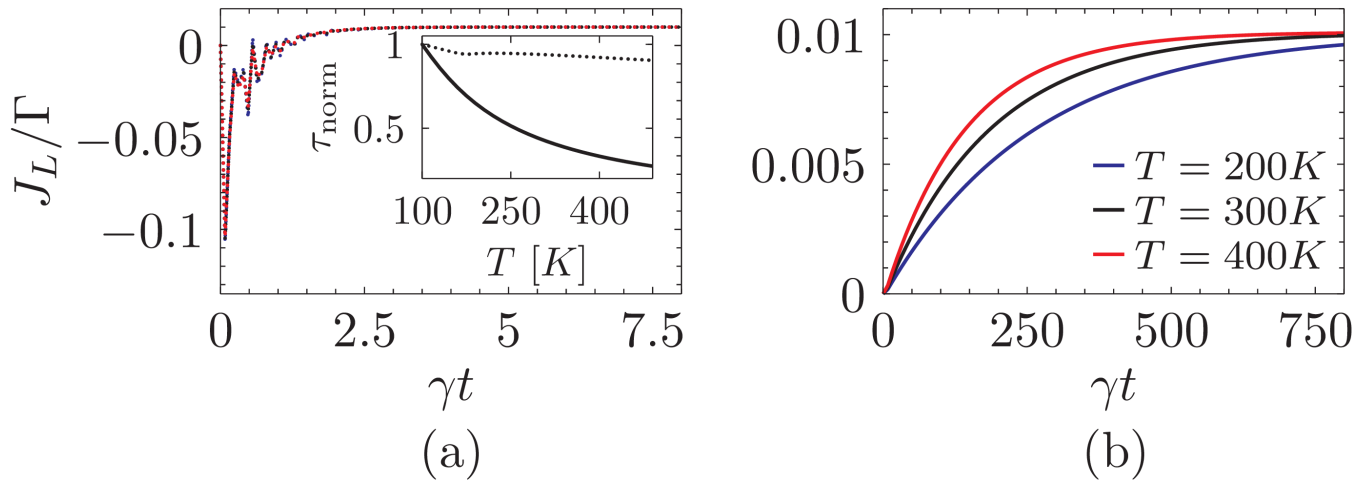}
\caption{\label{fig:figure3}\textbf{Probe temperature $T$ dependence on the dynamics of excitonic currents.} (a) shows the current as a function of time for different temperatures [$T=200 K$ (blue curve), $300 K$ (black curve), and $400K$ (red curve)]  when the cutoff frequency of the probe is very large $\omega_{D} = 58.80$ PHz, i.e.  $\tau_{ P} \ll \tau_{ S}$, for any temperature the behavior is the same and all the lines collapse, whereas (b) depicts the small cutoff frequency case $\omega_D = 1.96$ THz, i.e.,  $\tau_{ P} \gg \tau_{ S}$. Inset in panel (a) depicts the normalized relaxation time $\tau_{\rm norm} = \tau_{ R | T}/\tau_{ R|T_{0}}$ where $\tau_{R|T}$ and $\tau_{R|T_0}$ are the relaxation times evaluated at temperature $T$ and $T_{0}=100$K. The solid curve in the inset is for the case when $\tau_{ P} \gg \tau_{ S}$, whereas the dotted line is for $\tau_{ P} \ll \tau_{ S}$. In both panels the system is initiated in an antisymmetric state. All other parameters are the same as in Fig.~\ref{fig:figure1}.}
\end{figure*}

The canonical initial condition has overlap with both the stable and unstable steady states. The unstable steady state does not contribute to the dynamics because its \emph{relative} overlap with the canonical initial condition is much smaller than the overlap with the stable steady state. Hence, the mechanism leading to dynamics that is independent of the probe position is very similar to the symmetric initial condition case. Thus, the canonical initial condition, that is extensively used to study transport using nonequilibrium Greens function \cite{Wang2014}, hierarchy equation of motion \cite{Chen2014}, and polaron transformation \cite{Wang2015}, makes it impossible to detect molecular symmetries.

Our results described above are robust even in presence of a weak disorder as shown by closed circles in Fig.~\ref{fig:figure1}c. If the disorder is weak (much weaker than the effect of the probe, i.e., $\delta^{2} \ll \gamma$), then the eigenspectrum of the dissipative Liouvillian (Fig.~\ref{fig:figure2} asterisk), except the eigenvalue corresponding to the steady state, is not affected by the disorder. Since the weak disorder acts on the longest timescale $\tau_{\rm dis} \propto \delta^{-2}$ of the problem, we observe the dynamics at an intermediate timescale by ignoring the longer times that are affected by the disorder. In this intermediate timescale regime the disorder is insignificant and the dynamics is similar to the \emph{perfectly} symmetric scenario (black solid and green dashed-dotted line in Fig.~\ref{fig:figure1}c). Hence, \emph{a priori} if it is known that a weak disorder affects the system of interest, its effects could be neglected by observing the current dynamics at an intermediate timescale.

Next we investigate the effects of probe properties on the dynamics when initiated from the antisymmetric state and the probe is placed at site $2$. Due to the broken symmetry the unstable steady state forms a Hermitian decay mode, i.e., right eigenoperator of the dissipative Liouvillian with non-zero real eigenvalue \cite{Prosen2013}, that remain closest to the zero eigenvalue (unique steady state). Since this mode is formed due to the presence of the probe its eigenvalue is sensitive to the probe properties thus influencing the relaxation time \cite{Znidaric2015}.

Although this argument seems straightforward there is a caveat in terms of the dominant timescales of the problem. The largest frequency modes of the probe corresponding to the cut-off frequency $\omega_D$, Eq.~(\ref{eq:spec}), dictate a timescale $\tau_{P} \propto \omega_{D}^{-1}$ \cite{Weiss2012}, herein termed as the probe timescale. On the other hand the coherent system dynamics timescale is given by $\tau_{S} \propto \Delta$ where $\Delta$ is the smallest finite energy difference in the system eigenspectrum. These timescales play an influential role in determining the relaxation time of the system $\tau_{R}$. Figure~\ref{fig:figure3} shows the effect of interplay between these two timescales. When the system dynamics timescale dominates over the probe $\tau_{ S} \gg \tau_{ P}$, the probe has very little influence on the relaxation time $\tau_{M}$. The probe merely creates a unique steady state and a change in probe parameters, e.g. temperature, does not strongly influence the eigenvalue of the unstable steady state $\Lambda_{1}$. Thus, as a result the system relaxes quickly to the unique steady state and a change in probe temperature shows no effect as seen in the inset of Fig.~\ref{fig:figure3}a. On the other hand when the probe timescale dominates the system $\tau_{ P} \gg \tau_{ S}$ the probe can strongly influence the relaxation time $\tau_{R}$. In this case a change in the parameters, e.g. probe temperature, causes a proportional shift of the unstable steady state along the real axis thus influencing the relaxation time (inset of Fig.~\ref{fig:figure3}a). Since the shift is proportional to the temperature of the probe it is expected that the eigenvalue $\Lambda_{1}$ would be very close to zero for small values of temperature. In other words when $\tau_{ P} \gg \tau_{ S}$ the reorganization energy, $\gamma_{\rm reorg} = \int_{0}^{\infty} J(\omega)/(\pi \omega)\equiv \gamma \omega_{D}$ with $J(\omega)$ being the spectral density defined in Eq.~(\ref{eq:spec}), is extremely small and the probe acts like a weak perturbation to the system, whereas when $\tau_{ S} \gg \tau_{ P}$ the probe acts as a strong perturbation wiping out the symmetry information. Therefore, initiating from the antisymmetric state at low temperatures leads to a slow quasi-stationary relaxation towards the unique nonequilibrium steady state due to the strong interaction of the initial condition with the unstable steady state.

Thus, overall by changing the probe position and carefully engineering the initial state to a dark state \cite{Manzano2014, Daniel2015, Prosen2013} we detect the signatures of molecular symmetries. These signatures correspond to observing \emph{exactly} zero excitonic currents for symmetric systems and a finite time-dependent current with a long quasi-stationary plateau for systems with broken symmetry via the probe. Interestingly, the signatures completely vanish for the symmetric and canonical initial conditions making these an unacceptable choice for symmetry detection.

\begin{figure*}
\includegraphics[width=\textwidth]{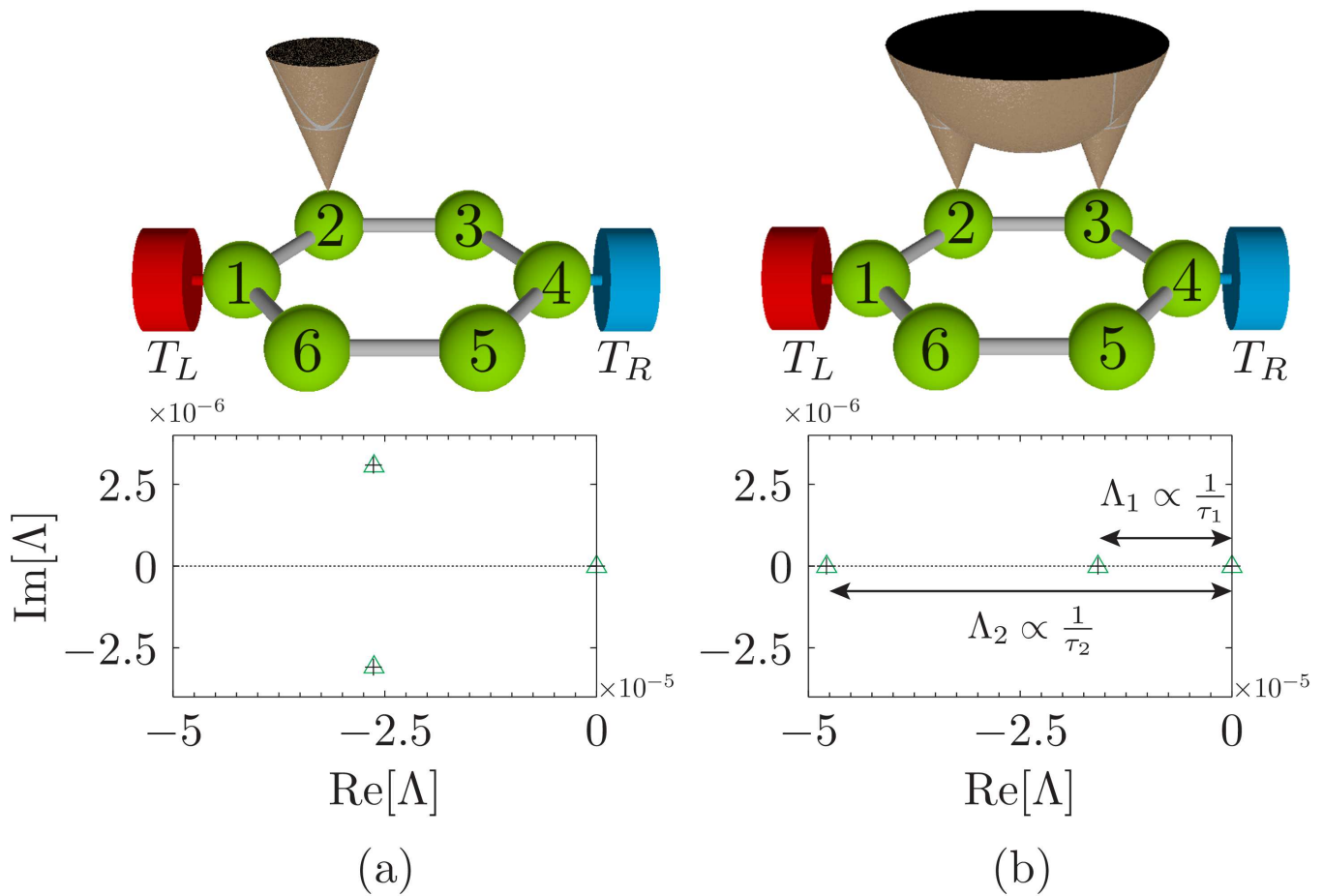}
\caption{\label{fig:figure4}\textbf{Eigenvalues of the dissipative Liouvillian for the para-benzene model.} (a) Eigenvalues closest to zero when the para-benzene symmetry is broken via a local probe acting on site $2$ (as depicted in the illustration above). (b) the probe is acting non-locally on sites $2$ and $3$ (illustrated above panel b) to break all the symmetries of the system. The black plus corresponds to the eigenvalues of the perturbation matrix $\delta L$ [Eq.~(\ref{eq:pertM})], whereas the green open triangles correspond to the smallest (in magnitude) eigenvalues of the full dissipative Liouvillian. The dashed black line marks the Im[$\Lambda$] = 0 axis. The system parameters are: $\varepsilon = -11.2$eV and $h = -0.7$eV. The lead and probe parameters are chosen as: $T_{ L} = 330K$, $T_{ R} = 270K$, $T_{ P} = 300K$, $\Gamma = 151.9$THz, $\gamma = 15.19$THz, $\omega_D = 151.92$THz and $\omega_0 = 78.55$THz.}
\end{figure*}

\noindent \textbf{Benzene molecule.} The H\"{u}ckle theory Hamiltonian $H_{ S}^{\prime}$ for benzene is given by
\begin{align}
H_{{ S}}^{\prime} &= \varepsilon \sum_{i=1}^{6}\op{e_i}{e_i} + h \sum_{\langle i,j \rangle} \op{e_i}{e_j}  
\end{align} 
wherein we choose the parameters $\varepsilon = -11.2$eV and $h_1 = -0.7$eV \cite{Salem1972}. The benzene molecule is connected to Lindblad leads similar to Eqs.~(\ref{eq:Lindblad1}) and (\ref{eq:Lindblad2}). The symmetry operator for the para-benzene ring $\Pi=\exp [\pare{\op{2}{6}+\op{6}{2}}\otimes\pare{\op{3}{5}+\op{5}{3}}]$ due to which the system has $3$ multiple steady states. Only one of these multiple steady states contains nonequilibrium information (e.g. temperature $T_{L}, T_{R}$ or coupling strength $\Gamma$) about the leads. The other two are pure states are decoupled from the leads and depend only on the symmetries. 

The steady states that depend only on the symmetries can be obtained analytically and they are given by
\begin{align}
\rho_{1} &= \op{\psi_1}{\psi_1} ,\nonumber \\
\ket{\psi_{1}} & =\frac{1}{2} \left( \ket{e_5} + \ket{e_6} - \ket{e_2} - \ket{e_3} \right), \\
\rho_{2} &= \op{\psi_2}{\psi_2} ,\nonumber \\
\ket{\psi_{2}} & =\frac{1}{2} \left( \ket{e_3} + \ket{e_6} - \ket{e_2} - \ket{e_5} \right).
\end{align}
wherein $\ket{\psi_i}$ are eigenvectors of the benzene Hamiltonian. The nonequilibrium steady state $\rho_{0}$ depends on all the lead properties and is nontrivial to obtain. In the presence of a probe $\rho_1$ and $\rho_2$ become unstable. In order to understand how the probe affects the multiple steady states we project the full dissipative Liouvillian, Eq.~(\ref{eq:ME}), into the subspace of the multiple steady states. This is achieved by transforming the steady states into vectors and the dissipative Liouvillian into a matrix as done in Eq.~(\ref{eq:rhot}). Thus, the $3\times 3$ perturbation matrix reads
\begin{align}
\delta L &=\left( \begin{array}{ccc}
\theta_0& \sigma_{1} & \sigma_{2}  \\
\theta_1& R_{11} & R_{12}  \\
\theta_2& R_{21} & R_{22}  \end{array} \right)
\label{eq:pertM}
\end{align}
where the elements $\theta_{n}$ ($n=0,1, 2$) contain information about the nonequilibrium steady state and the probes and $\sigma_{k} \propto \sum_{j\neq1,2}R_{jk}$. The matrix elements $R_{jk}$ are the elements of the Redfield tensor $R[\rho] = \int_{0}^{\infty}dt [S,\rho S(t)]C(t)+\mathrm{h.c.}$ (see methods section Eq.~(\ref{eq:ME}) and also Refs.~\cite{Cao1997, Thingna2012} for more details) and are given by
\begin{align}
R_{jk} &= \sum_{p}2\left\{S_{jk}^{p}S_{kj}^{p}\mathrm{Re}[W_{jk}^{p}] -\delta_{j,k}\sum_{l}S_{jl}^{p}S_{lk}^{p}\mathrm{Re}[W_{lk}^{p}]\right\}.
\end{align}
Above we have taken into consideration that several independent probes could act on the system whose actions are taken into account via the summed index $p$. The operator elements are $S_{jk} = \bra{\psi_j}S\ket{\psi_k}$ and $W_{jk}= \int_{0}^{\infty}dt e^{-i\Delta_{jk}t}C(t)$ with $\Delta_{jk}=E_j-E_k$. Here $E_k$ is the eigenenergy of the benzene Hamiltonian corresponding to the state $\ket{\psi_k}$. The perturbation matrix $\delta L$ has one zero eigenvalue corresponding to the unique nonequilibrium steady state and two non-zero eigenvalues that dictate the relaxation time. The structure of the matrix $\delta L$ clearly indicates that the elements $R_{jk}$ govern the effect of the probe on the symmetries of the system. Since the probe obeys detailed balance the backward transition rates $\mathrm{Re}[W_{12}^{p}]$ are extremely small. Thus, we can neglect the matrix element $R_{12}\approx 0$ and simplify the perturbation matrix. This simplification occurs only due to the probe properties and does not depend on how the probe connects to the system.
\begin{figure}
\includegraphics[width=\columnwidth]{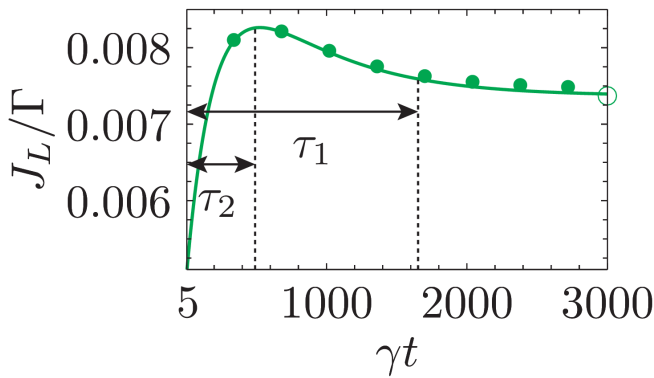}
\caption{\label{fig:figure5}\textbf{Bi-exponential decay of currents as a signature of $3$ steady states.} Long-time dynamics of the para-benzene model exhibiting a bi-exponential decay, due to which a peak is formed, when the system is initiated in an antisymmetric state w.r.t. sites $2$ and $5$. The probe acts non-locally on sites $2$ and $3$ as illustrated above panel (b) in Fig.~\ref{fig:figure4}. The open green circle corresponds to the steady state value for the current. The closed green circles are for a weakly disordered system with disorder strength $\delta = 1$meV. The currents are averaged over 1000 samples of uniformly distributed random disorder [standard error of mean (SEM) is contained within each closed circle]. All other parameters are same as Fig.~\ref{fig:figure4}.}
\end{figure}

The matrix element $R_{21}$ controls the structure of the eigenvalues of the perturbation matrix and depends only on how the probe connects to the system (via operator $S$). In particular, we find that when $R_{21} \neq 0$ the non-zero eigenvalues of $\delta L$ form a complex conjugate pair. On the other hand when $R_{21} = 0$ we break all the symmetries and obtain three distinct eigenvalues of the perturbation matrix. In order to validate our claim we take the simplest example of a single probe $p=1$ that breaks the symmetry of the system and connects via a general operator $S =\sum_{l=g,1}^{6}C_l\op{e_l}{e_l}$ with $C_l$ being constants. Thus, in order to achieve $R_{21} = 0$ we require
\begin{align}
S_{12} = \frac{1}{4}\left(C_2-C_3-C_5+C_6\right)=0.
\label{eq:cond}
\end{align}
If $C_2=C_3=C_5=C_6$ then the probe is affecting all the symmetric sites of benzene equally without breaking any symmetry. Thus, even though Eq.~(\ref{eq:cond}) is satisfied we obtain $3$ multiple steady-states. Hence it is of utmost importance that the probe breaks at least one of the symmetries of benzene. If we choose a local probe acting on site $2$, i.e., $C_3=C_5=C_6 =0$, it breaks some but not all the symmetries [since Eq.~(\ref{eq:cond}) is not satisfied]. This case leads to a unique steady state even if not all the symmetries are broken. The eigenvalues of the perturbation matrix (black plus) and the full dissipative Liouvillian (green triangles) are plotted in Fig.~\ref{fig:figure4}a for this case. Due to the fact that we are not breaking all the symmetries, we obtain a couple of non-zero eigenvalues that are conjugated and there is only one relaxation time (given by the real part of the non-zero eigenvalue). The dynamical currents are unable to show any \emph{exclusive} signatures of the three steady states. The current dynamics are qualitatively similar to the $4$-site model (see supplementary material). In order to break \emph{all} the symmetries we take the simplest case $C_2=C_3$ and $C_5=C_6=0$. In this case the probe acts non locally on sites $2$ and $3$ [Eq.~(\ref{eq:cond}) is satisfied]. This leads to three distinct real eigenvalues of the perturbation matrix as shown in Fig.~\ref{fig:figure4}b.

The left-eigenvectors corresponding to the distinct non-zero eigenvalues inherit the symmetry properties of the unstable steady states. Particularly the left-eigenvector corresponding to the eigenvalue closest to zero retains the symmetry of $\rho_1$ whereas the one further away transforms into a linear combination of $\rho_1$ and $\rho_2$. Thus, we initiate the system in a state that is antisymmetric in sites $2$ and $5$ (equivalent to $\rho_1$) and find that at long times both the non-zero eigenvalues closest to the steady state ($\Lambda_1$ and $\Lambda_2$) will contribute to the relaxation time. Hence, due to the remnants of the unstable steady states in the Hermitian decay modes we can uncover the signatures of the multiple steady state as seen in the current dynamics in Fig.~\ref{fig:figure5}. At long times the current dynamics exhibits two distinct exponential relaxation times $\tau_1 \propto \Lambda_1^{-1}$ and $\tau_2 \propto \Lambda_2^{-1}$. The corresponding coefficients of the exponential decays are are opposite in sign and hence the excitonic current shows a peak indicating the presence of two unstable steady states. Here once again similar to the $4$-site model a weak static disorder in the system, as shown by the closed green circles in Fig.~\ref{fig:figure5}, plays no role at the relevant timescale. Thus, by redesigning the probe to act non locally it is possible to detect the number of multiple steady states by observing the multi-exponential decay at long times.\\
~\\
\noindent \textbf{Discussion and Conclusions}\\
\indent Several molecular systems possess symmetries that could play a crucial role in the nonequilibrium properties of the quantum devices. Naturally, one wonders if it is possible to detect symmetries without any prior knowledge of the molecular system. In the present study, we provide a toolbox that detects molecular symmetries via the measurement of excitonic currents. 

Our goal is achieved by introducing a probe that acts on various sites of the molecular junction and depending on the probe position the currents differ suggesting the presence of symmetries. In particular, we find signatures of molecular symmetries if we start from an antisymmetric (dark) state as our initial condition. If the probe is then positioned on a site that preserves the molecular symmetry we find zero excitonic currents because the dark state belongs to an invariant subspace of the nonequilibrium set-up. On the other hand if the probe is placed on one of the symmetric sites it breaks the inherent symmetry of the system leading to a quasi-stationary plateau in the excitonic currents. When the probe is placed on any of the symmetric sites we see identical time-dependence of the excitonic current indicating that these sites are interchangeable, i.e., symmetric. These signatures vanish if we initiate the system in the symmetric or canonical initial condition making the antisymmetric state a primary choice for symmetry detection.

Our detection scheme is based on dynamical signatures in transient currents and is robust against the weak disorder. We specifically show that in the presence of disorder the steady-state currents do not show any symmetry related signatures, whereas the transients exhibit dynamical signatures related to the the perfectly symmetric systems. The strength of the disorder must be weaker than that of the probe, such that the effect of disorder does not appear except for the longest timescale. Thus, by ignoring the longest timescale $\tau_{{\rm dis}}\propto \delta^{-2}$ the dynamics at intermediate timescales becomes identical to that without disorder. In other words, even for a weakly disordered symmetric system we could detect the underlying symmetries using the probe, since the dynamics at intermediate timescale mimics the long-time dynamics of the perfectly symmetric system.

Since the antisymmetric state is our prime choice for symmetry detection we also study the effect of probe properties on the dynamics, if we initiate the system from this state. We find that  the interplay between the timescales of the probe and the system dynamics play a significant role in determining the dynamics and the relaxation time of the system. In particular, when the molecular symmetry is broken, the system possess a long relaxation time that scales with temperature if the timescale of the probe dynamics is much larger than the system ($\tau_{ P} \gg \tau_{ S}$). In the opposite case temperature shows no effect on the relaxation time indicating that the probe does not affect the unstable steady state leading to quick relaxation. The long relaxation time indicates the presence of a long-lived quantum quasi-stationary distribution \cite{Macieszczak2015} that survives due to its interaction with the unstable steady state.

In case of the symmetric para-benzene molecule we found three nonequilibrium steady states. In order to break all the symmetries we analyzed the system by treating the probe as a perturbation and showed that the multiple symmetries can be broken via a redesigned probe. Our new probe acted on two sites of the para-benzene ring simultaneously and led to a bi-exponential decay in the current dynamics. The bi-exponential decay exhibited long relaxation times and the exponents corresponded to the eigenvalues of the unstable manifold. This trend of multi-exponent decays corresponding to the number of symmetries present was then recognized as the second key signature to detect symmetries.

Overall, our work provides a first glimpse into symmetry detection in nonequilibrium transport set-ups and could be extended to systems such as conjugated dendrimers \cite{Wu2006}, light-harvesting complexes \cite{Cleary2013}, C60 bucky balls \cite{Geranton2013}, etc. to detect naturally occurring symmetries. Also since quantum coherence plays  a key role in the performance of light-harvesting systems and quantum heat engines \cite{Gelbwaser2015, Xu2016}, it naturally leads to the speculation that structural symmetry studied here will lead to interesting quantum effects in these systems
\\
~\\
\noindent \textbf{Methods}\\
\noindent \textbf{Invariant subspaces.}
It has been previously proved that a Lindblad master equation has invariant subspaces if there is a unitary operator that commutes with all the master equation elements \cite{Buca2012, Albert2014}. In this section we extend this proof to a general open system. We begin with a microscopic description and decompose the total Hamiltonian as $H = H_{{ S}} + H_{{B}} + H_{{ BS}}$. The Hilbert space of the total system is decomposed in the direct product of our system of interest $({ S})$, and the bath $({ B})$, $\mathcal{H}=\mathcal{H}^{ S}\otimes\mathcal{H}^{ B}$ with $H_{ S}\in O\left( \mathcal{H}^{ S}\right)$, $H_{ B} \in O \left(\mathcal{H}^{ B}\right)$, and $H_{ BS}= \sum_{j}S \otimes Y_{j}\in O\left(\mathcal{H}^{ S}\right) \otimes O\left(\mathcal{H}^{ B}\right)$. The bath $B$ in our description could comprise of several leads and probes, e.g., $H_{B} = H_{L} + H_{R} + H_{P}$, where the sub-scripts $L$, $R$, and $P$ correspond to the left-lead, right-lead, and the probe. 

We assume that there is a unitary operator $\Pi\in O\left(\mathcal{H}^{ S}\right)$ s.t. $[ \Pi,H_{ S}]=[\Pi,S]=0$. The total Hilbert space can be decomposed in mutually orthogonal eigenspaces of $\Pi$, i.e., $\mathcal{H}=\oplus_{i=1}^{n_s} \mathcal{H}_i $ with $n_s$ being the number of eigenvalues of $\Pi$. In an equivalent way the  Hamiltonian of the system can be decomposed $\mathcal{H}_{ S}=\oplus_{i=1}^{n_s} \mathcal{H}^{ S}_i $. As the operator $\Pi$ is compatible with all the components of the Hamiltonian they share an eigenbasis and each part of the Hamiltonian can be block-diagonalised. This implies that if $|\psi_\alpha\rangle \in \mathcal{H}_{k}$  the following statements hold 
\begin{align}
H_{ S}|\psi_{k}\rangle \in \mathcal{H}_{k};~H_{ BS}|\psi_k\rangle \in \mathcal{H}_k;~S|\psi_k\rangle \in \mathcal{H}_k.
\label{eq:hspectral}
\end{align}
To deal with density matrices we define the adjoint operator $\tilde{V}\in O \left(\mathcal{H}\right)$ as $\tilde{V}(\varrho)=V\varrho V^\dagger$ ($\varrho$ is the density matrix for the total Hamiltonian). If the eigenvalues of $V$ are $\nu_k=e^{i\Omega_k},\; (k=1,2,...,n_s)$, the adjoint operator eigenvalues are just the product of these eigenvalues $\nu_k \nu_l=e^{i\left(\Omega_k-\Omega_l\right)}$. The adjoint space $O \left(\mathcal{H}\right)$ can be decomposed as $O\left(\mathcal{H}\right)=\oplus_{k=1}^{n_s} \oplus_{l=1}^{n_s} V_{kl}$. If a density matrix has the form $\varrho_{kl}=|\psi_{k}\rangle\langle\psi_{l}|$, it means that $\varrho_{kl}\in V_{kl}$. From the spectral decomposition, together with Eq. (\ref{eq:hspectral}), we can derive that if $\varrho_{kl}\in V_{kl}$ then 
\begin{align}
[H_{ S},\varrho_{kl}] \in V_{kl};~[H_{ B},\varrho_{kl}]\in V_{kl};~[S,\varrho_{kl}]\in V_{kl}.
\label{eq:com}
\end{align}

We work in the interaction picture, where the dynamics of the system is given by the interaction Hamiltonian, $\dot{\varrho}(t)=-i[H_{ BS}(t),\varrho(t)]$, and the time dependence of the interaction Hamiltonian is given by
\begin{align}
H_{ BS}(t) &= e^{i\left(H_{ S}+H_{ P}\right)t}H_{ BS} e^{-i\left(H_{ S}+H_{ B}\right)t}.
\label{eq:interaction}
\end{align}
Combining Eqs.~(\ref{eq:com}) and (\ref{eq:interaction}) it can be proved that 
\begin{align}
[H_{ BS}(t),\rho_{kl}] \in V_{kl}.
\label{eq:comm}
\end{align}
It is clear from Eq.~(\ref{eq:comm}) and the von Neumann equation that if $\varrho(0)\in V_{kl}$ then $\varrho(t) \in V_{kl}$. Consequently, since tracing over the probe degrees of freedom doesn't cause a change in subspaces $\implies \rho(t) \in V_{kl}$. This completes our general proof of the invariant subspaces in open quantum systems.

\noindent \textbf{Redfield-Lindblad quantum master equation.}
The nonequilibrium transport set-up described within this work can be modeled using the total Hamiltonian,
\begin{align}
H&=H_{{ S}} + H_{{ L}} + H_{{ R}} + H_{{ P}} + H_{{ LS}} + H_{{ RS}} + H_{{ PS}},
\end{align}
wherein $H_{{ S}}$ describes the system and $H_{\alpha}$ describes the left-lead, right-lead and the probe with $\alpha =$ L, R and P respectively. The interaction of the system with the leads and the probe is given via $H_{\alpha {{ S}}}$. The leads and the probe Hamiltonian are given by an infinite set of independent harmonic oscillators that read
\begin{align}
H_{\alpha} &=\frac{1}{2}\sum_k \frac{p_k^{\alpha\,2}}{m_k^{\alpha}}+m_{k}^{\alpha}\left(\omega_k^{\alpha} x_k^{\alpha}\right)^{2}.
\label{eq:Hamiltonians}
\end{align}
We assume the system-lead/probe interaction to be weak (Born-approximation) and the leads/probe to be memory-less (Markov-approximation). Additionally, only for the leads, we assume that the dissipative effects of the leads should be relevant on a timescale much longer than all finite times of the problem (secular-approximation). Under these assumptions and initiating the total system in a product state of the system, leads, and probes, with no cross-correlations between the leads and the probe, the master equation for the reduced density matrix reads \cite{Alicki2007, Breuer2007},
\begin{align}
\label{eq:ME}
\dot{\rho} =& \mathcal{L}\rho \nonumber \\
\mathcal{L}\rho =& - i [H_{{ S}},\rho] + \sum_{\substack{k=1,2\\ \alpha={ L, R}} } \Gamma_{\alpha k} \left( A_{\alpha k}^{\phantom{\dagger}}\rho A_{\alpha k}^{\dagger}-\frac{1}{2}\{A_{\alpha k}^{\dagger}A_{\alpha k}^{\phantom{\dagger}},\rho\}\right) \nonumber \\
& +\int_{0}^{\infty}dt \left\{ [S,\rho S(t)]C(t)+\mathrm{h.c.} \right\}.
\end{align}
Throughout this work we set $\hbar =1$ for notational simplicity. The secular approximation causes the effect of the leads on the system to be described via the Gorini-Kossakowski-Sudarshan-Lindblad master equation \cite{Gorini1976, Lindblad1976}. The Lindblad operators $A_{\alpha k}$ corresponding to the $\alpha$-th lead allow local injection ($k=1$) or extraction ($k=2$) of one exciton to the molecular system $H_{{ S}}^{\prime}$ (see Eq.~(\ref{eq:H})). The rates $\Gamma_{\alpha i}$ dictate the temperature of each lead, namely, $\Gamma_{\alpha 1} = \Gamma n_{\alpha} $ and $\Gamma_{\alpha 2} = \Gamma (n_{\alpha}+1)$ with $\Gamma$ being the dissipation strength of each lead and $n_{\alpha} = [\exp(\beta_{\alpha}\omega_0)-1]^{-1}$ the Bose-Einstein distribution function. The ratio of the rates $\Gamma_{\alpha 1}/\Gamma_{\alpha 2} = \exp (-\beta_{\alpha}\omega_{0})$ obeys detailed balance where $\beta_{\alpha}=1/k_{{\rm B}}T_{\alpha}$ represents the inverse temperature of the $\alpha$-th lead and $\omega_0$ is the characteristic phonon frequency of the lead.

The effect of the probe on the system is described via the Redfield tensor \cite{Redfield1957} that allows us to study the interplay of various additional timescales (due to the absence of secular approximation), e.g. the system and the probe dynamics timescales. The operator $S$ in the Redfield tensor originates from the system-probe interaction Hamiltonian that we choose to be of a general form $H_{{ PS}} = S\otimes Y$. The operator $Y$ is then encapsulated in the correlator $C(t)={\rm Tr}_{P}\left[\tilde{Y}(t)Y(0)\exp\left(\beta H_{{ P}}\right)\right]$ where $\beta$ is the inverse temperature of the probe and $\tilde{Y}(t)$ is the free-evolution of the probe operator $Y$ with respect to the probe Hamiltonian. We choose the probe operator to be the collective position operator $Y=-\sum_{k}c_{k} x_{k}$ with $c_k$ being the coupling strength of each harmonic mode of the probe to the system. Throughout this work, the explicit super-script ${ P}$ will be suppressed for convenience. All parameters of the probe are described via a spectral density
\begin{align}
J(\omega) &= \pi \sum_{k=1}^{\infty}\frac{c_{k}^{2}}{2m_{k}\omega_{k}} \delta(\omega-\omega_{k}) =\frac{ \gamma\omega}{1+(\omega/\omega_D)^2},
\label{eq:spec}
\end{align}
that is chosen to be of the ohmic form with a Lorentz-Drude cut-off frequency $\omega_D$ and dissipation strength $\gamma$. The corresponding correlator $C(t)$ is given by,
\begin{align}
C(t) &=\int_{0}^{\infty}\frac{d\omega}{\pi}J(\omega)\left[{\rm coth}\left(\frac{\beta\omega}{2}\right)\cos(\omega t) - i\sin(\omega t)\right].\nonumber
\end{align}

\noindent \textbf{Full counting statistic for excitonic transport.}
In order to quantify the excitation flux through the system we include a counting field in the quantum master Eq.~(\ref{eq:ME}) \cite{Manzano2014, Esposito2009, Garrahan2010}. We first introduce the reduced density matrix $\rho_{q}(t)$ that is the projection of the density matrix into the subspace of $q$ excitations interchanged between the system and the left-lead at time $t$. The probability of observing an exciton current $J_{q}$ is thus given by $P(q)=\mathrm{Tr} \left[\rho_q(t)\right]$. The evaluation of this probability is easier after a change of ensemble, made by introducing the Laplace transform $\rho_{\lambda}(t)= \sum_q \rho_q(t) \exp(-\lambda q)$ with $\lambda$ being a counting field. The evolution of the density matrix in the Laplace transformed ensemble $\rho_{\lambda}(t)$ is given by

\begin{align}
\dot{\rho}_{\lambda} =& - i [H_{ S},\rho_{\lambda}] +\int_{0}^{\infty}dt \left\{[S,\rho S(t)]C(t)+\mathrm{h.c.}\right\} \nonumber \\
&+ \sum_{k=1,2} \Gamma_{{ L} k} \left( e^{\left[(-1)^{k}\lambda\right]}A_{{ L} k}^{\phantom{\dagger}}  \rho A_{{ L} k}^{\dagger}-\frac{1}{2}\{A_{{ L} k}^{\dagger}A_{{ L} k}^{\phantom{\dagger}},\rho\}\right) \nonumber \\ 
&+ \sum_{k=1,2} \Gamma_{{ R} k} \left( A^{\phantom{\dagger}}_{{ R} k}\rho A_{{ R} k}^{\dagger}-\frac{1}{2}\{A_{{ R} k}^{\dagger}A_{{ R} k}^{\phantom{\dagger}},\rho\}\right) \nonumber \\
=&\mathcal{L}_{\lambda} \rho_{\lambda}.
\end{align}
Given the above density matrix as a function of the counting field $\lambda$ the moment generation function can be calculated as $Z_{\lambda}(t)\equiv {\rm Tr}\left[\rho_{\lambda} (t)\right]$. Thus, the average exciton current flowing into the system from the left-lead $J_{{ L}}$ can be obtained from the generation function as $J_{{ L}}= (1/t) \left[\partial_{\lambda} \log Z_\lambda (t)\right]_{\lambda=0}$. Considering that $\rho_{\lambda}(t)=\exp \left(+t \mathcal{L}_\lambda\right) \rho(0)$ we obtain the left-lead expected current as,
\begin{align}
J_{{ L}}&=\left.\frac{{\rm Tr}\left[ \left( \partial_\lambda  \mathcal{L}_\lambda\right) \rho_{\lambda}(t)\right]}{{\rm Tr}\left[\rho_{\lambda} (t)\right]} \right\vert_{\lambda=0}  \\
&= \Gamma_{{ L}1}\mathrm{Tr}\left[A_{{ L}1}^{\dagger}A_{{ L}1}^{\phantom{\dagger}} \rho(t)\right]-\Gamma_{{ L}2}\mathrm{Tr}\left[A_{{ L}2}^{\dagger}A_{{ L}2}^{\phantom{\dagger}}\rho(t)\right]. \nonumber
\end{align}

\noindent \textbf{Acknowledgement}\\
J. T. acknowledges financial support by SMART, D.M. acknowledges financial support from the Junta de Andaluc\'ia and EU, Project TAHUB/II-148 (Program ANDALUC\'IA TALENT HUB 291780) and the MIT-SUTD program, and J. C. is supported by NSF (grant no. CHE-1112825) and SMART.
~\\
\noindent \textbf{Author contributions}\\
J.T. developed the master equation approach and the excitonic transport theory. D. M. provided a general microscopic proof for the invariant subspaces. J. T. carried out the numerical simulations. J. C. proposed the project and methods. All three authors analyzed the results and contributed equally towards the presentation of the manuscript.\\
~\\
\noindent \textbf{Additional information}\\
\noindent \textbf{Competing financial interests:} The authors declare no competing financial interests.

\pagebreak
\section*{Supplemental material for \\`Dynamical signatures of molecular symmetries in nonequilibrium quantum transport'}

\begin{figure*}
\includegraphics[width=0.9\textwidth]{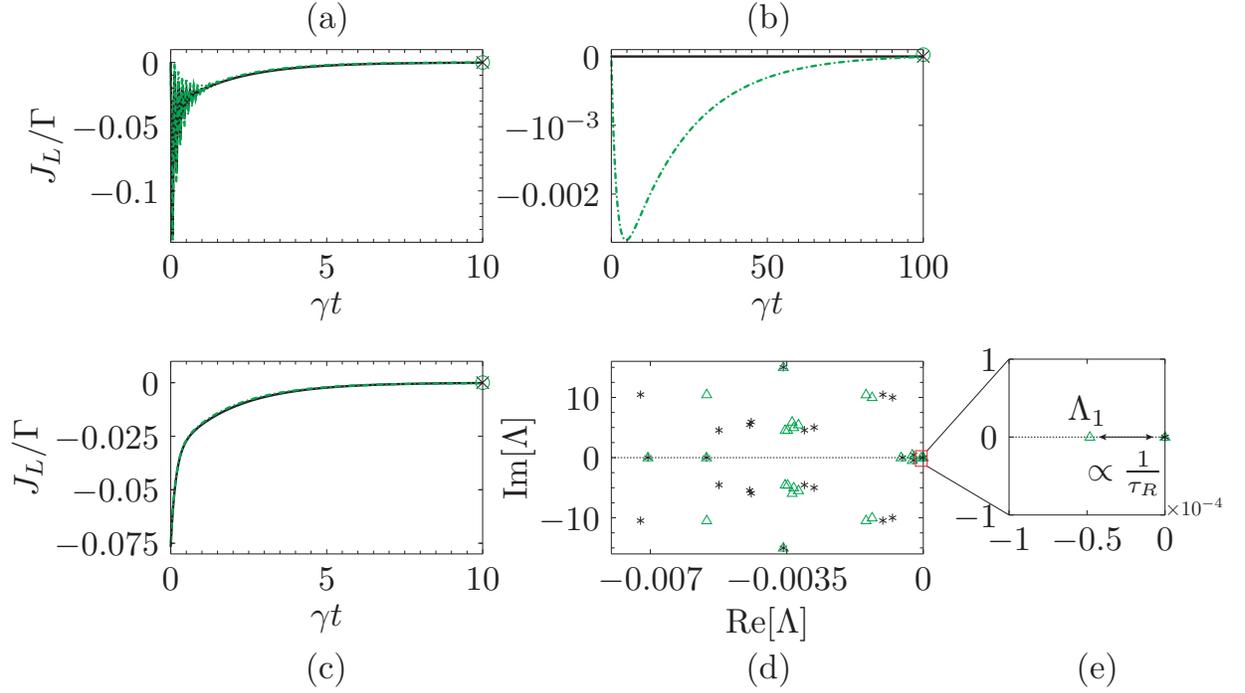}
\caption{\label{fig:figure1s}\textbf{Effect of probe position on symmetry detection.} Time evolution of the excitonic current in the $4$-site model with symmetric (panel a), antisymmetric (panel b), and canonical (panel c) initial conditions. (d) shows the eigenspectrum of the dissipative Liouvillian and (e) depicts the magnification around the zero eigenvalue. The dashed black line in panels (d) and (e) marks the Im[$\Lambda$] = 0 axis. The probe is positioned at sites $1$ (black solid line in panels a, b and c; black asterisk in panels d and e) and $2$ (green dashed-dotted line in panels a, b, and c; green triangles in panels d and e) for all panels. The crosses and circles correspond to the nonequilibrium steady-state values of the excitonic currents. The system parameters are: $\varepsilon_{1} = -64.6$meV, $\varepsilon_{3} =-193.9$ meV, $\varepsilon_{2}=\varepsilon_{4}=-129.3$ meV, $h_{12}=h_{14}=-14.52$ meV, and $h_{23}=h_{34}=-1.59$meV. The lead and probe parameters are chosen as: $T_{L} = 330K$, $T_{R} = 270K$, $T = 300K$, $\Gamma = 196$GHz, $\gamma =19.6$GHz, $\omega_0 = 78.55$THz, and $\omega_{D}=1.96$ THz.}
\end{figure*}
\noindent \textbf{$4$-site model revisited}\\ The results we show in the main text correspond to a system where diagonal and off-diagonal elements of the Hamiltonian $H_{S}^{\prime}$ are chosen to be the same. This requirement can be relaxed. Symmetries could exist even if these parameters are mismatched. In this section we explore a more general case with the help of the $4$-site model and show that the signatures of molecular symmetries are robust. Thus, if the system possess a molecular symmetry in any form our probe based approach could help detect these symmetries. We begin by choosing a general $4$-site Hamiltonian
\begin{align}
\label{eq:H4}
H_{{ S}}^{\prime} &= \sum_{i=1}^{4}\varepsilon_{i}|e_i\rangle\langle e_i| + \sum_{\langle i,j \rangle}h_{ij} |e_i\rangle\langle e_j|,
\end{align}
where the parameters $\varepsilon_{i}$ and $h_{ij}=h_{ji}$ are different for each site. In order to maintain the mirror symmetry of the system we choose $\varepsilon_{2}=\varepsilon_{4}$, $h_{12}=h_{14}$, and $h_{23}=h_{34}$.

Figure~\ref{fig:figure1s} shows the behavior of the excitonic currents when the probe is placed either at site $1$ (black solid lines) or at site $2$ (green dashed-dotted lines) for symmetric (panel a), antisymmetric (panel b) or canonical (panel c) initial conditions. The conclusions remain the same as in the main text, i.e., the signatures of molecular symmetries is observed only in the case of a dark state initial condition. One main difference we observe when all parameters are not the same is the change in the dissipative Liouvillian spectrum (Fig.~\ref{fig:figure1s}d). Contrary to the fully identical case (see Fig.~2 of main text) the eigenspectrum changes drastically. Although the splitting of the degenerate eigenvalue still persists and the unstable manifold still remains closest to the zero eigenvalue thus influencing the long relaxation times of the system.

\begin{figure*}
\includegraphics[width=0.9\textwidth]{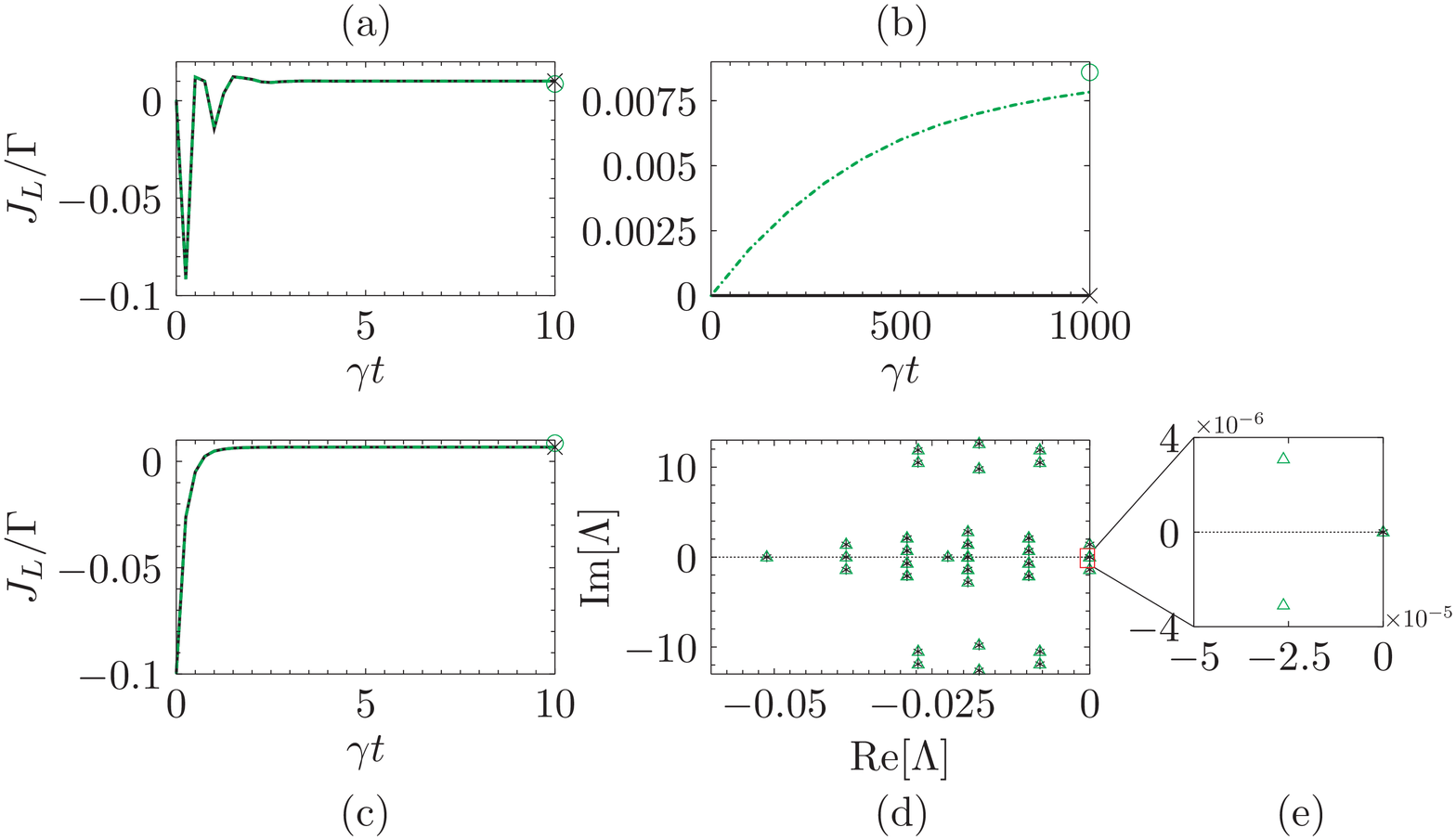}
\caption{\label{fig:figure2s}\textbf{Effect of probe position on symmetry detection.}Time evolution of the excitonic current for the Benzene molecule with symmetric in sites $2$ and $6$ (panel a), antisymmetric in sites $2$ and $6$ (panel b), and canonical (panel c) initial conditions. The crosses and circles correspond to the nonequilibrium steady-state values of the excitonic currents. Panel (d) shows the eigenspectrum of the dissipative Liouvillan and (e) depicts the magnification around zero eigenvalue. The dashed black line in panels (d) and (e) marks the Im[$\Lambda$] = 0 axis. The probe is positioned at sites $1$ (black solid line panels a, b and c; black asterix in panels d and e) and $2$ (green dashed-dotted line in panels a, b, and c; green triangles in panels d and e). The system parameters are: $\varepsilon = -11.2$eV and $h = -0.7$eV. The lead and probe parameters are chosen as: $T_{ L} = 330K$, $T_{ R} = 270K$, $T_{ P} = 300K$, $\Gamma = 151.9$THz, $\gamma = 15.19$THz, $\omega_D = 151.92$THz and $\omega_0 = 78.55$THz.}
\end{figure*}
\noindent \textbf{Effect of probe position in para-Benzene}\\ Similar to the $4$-site model shown in the main text the effect of a local probe on benzene is shown in Fig.~\ref{fig:figure2s}. The qualitative features remain the same as in the $4$-site system with the anti-symmetric state showing the clear signatures of multiple steady states. The Liouvillian spectrum also shows exotic modes like oscillating coherences \cite{Albert2014} that correspond to eigenvalues of the Liouvillian with zero real part but finite imaginary part. These could be the modes responsible for the observation of circular currents observed in benzene \cite{Rai2010}.

\end{document}